# Cavity-enhanced narrowband spectral filters using rare-earth ions doped in thin-film lithium niobate

Yuqi Zhao[1,2*], Dylan Renaud[3], Demitry Farfurnik[4], Yuxi Jiang[1,2], Subhojit Dutta[1,2], Neil Sinclair[3*], Marko Lončar[3], Edo Waks[1,2*]

[1]Department of Electrical and Computer Engineering, University of Maryland, College Park, MD, 20742, USA
[2]Institute for Research in Electronics and Applied Physics (IREAP), University of Maryland, College Park, MD, 20742, USA
[3]John A. Paulson School of Engineering and Applied Sciences, Harvard University, 29 Oxford Street, Cambridge, MA, 02138, USA
[4]Department of Electrical and Computer Engineering, and Department of Physics, North Carolina State University, Raleigh, NC, 27695, USA
*yuq1zhao@umd.edu, neils@seas.harvard.edu, edowaks@umd.edu

## Abstract

On-chip optical filters are fundamental components in optical signal processing. While rare-earth ion-doped crystals offer ultra-narrow optical filtering via spectral hole burning, their applications have primarily been limited to those using bulk crystals, restricting their utility. In this work, we demonstrate cavity-enhanced spectral filtering based on rare-earth ions in an integrated nonlinear optical platform. We incorporate rare-earth ions into high quality-factor ring resonators patterned in thin-film lithium niobate. By spectral hole burning at 4K in a critically coupled resonance mode, we achieve bandpass filters ranging from 7 MHz linewidth, with 13.0 dB of extinction, to 24 MHz linewidth, with 20.4 dB of extinction. By reducing the temperature to 100 mK to eliminate phonon broadening, we achieve an even narrower linewidth of 681 kHz, which is comparable to the narrowest filter linewidth demonstrated in an integrated photonic device, while only requiring a small device footprint. Moreover, the cavity enables reconfigurable filtering by varying the cavity coupling rate. For instance, as opposed to the bandpass filter, we demonstrate a bandstop filter utilizing an under-coupled ring resonator. Such versatile integrated spectral filters with high extinction ratio and narrow linewidth could serve as fundamental components for optical signal processing and optical memories on-a-chip.

## Introduction

On-chip optical filters are a fundamental building block in optical computing[1], optical spectrometry[2], sensing[3], microwave photonics[4], ultrafast pulse shaping[5–7], and wavelength division multiplexing[8]. Specifically, optical filters in the MHz and kHz range are crucial to many classical applications in biomedical imaging[9,10], radio-astronomy[11,12], remote sensing[13], and the processing of radio-frequency, microwave, or millimeter-wave signals using photonics[14–16]. They also play an essential role in quantum information applications such as filtering of narrowband single photon sources[17–19], reduction of background noise in quantum frequency conversion[20,21], and single photon pulse shaping[22]. However, integrated optical filters that can simultaneously achieve narrow bandwidth, low loss, and high extinction ratios are lacking, particularly for filter bandwidths below tens of MHz[23].

Rare-earth-ion-doped crystals can realize ultra-narrow optical filters via spectral hole burning[23]. In this approach, a narrowband laser excites the ions to remove a narrow absorption line from the inhomogeneously broadened absorption spectrum of the ions, resulting in a bandpass filter at the targeted wavelength. The resulting linewidth could approach two times the ion homogeneous linewidth, which could be very narrow, e.g., 146 Hz in Er:YSO[24]. However, such optical filters have primarily been studied in bulk crystals, with motivation primarily for ultrasound detection[16,25,26], radio-frequency spectrum analysis[14,27] and spectral-spatial holography[28]. While rare-earth-ion-doped nanophotonics have been extensively explored for quantum memory[29–31], single ion qubit[32], and cavity quantum emitters[33], on-chip optical filtering and signal processing using rare-earth ions has not been thoroughly investigated.

Thin-film lithium niobate has emerged as a promising integrated host platform for rare-earth ions. Furthermore, this material has underpinned a range of high-performance integrated nanophotonic components[34] due to its exceptional electro-optic and nonlinear properties[35]. Rare-earth ion based amplifiers[36,37], lasers[38], as well as quantum memories[31] have been demonstrated in thin-film lithium niobate. The ultra-low propagation loss (3 and 6 dB/m at 1550 nm[39,40] and 638 nm[41] respectively) in this platform has also enabled high quality factor photonic cavities, which has already achieved Purcell enhancement of single ion emission[42] and long range cooperative resonances[33]. These distinctive properties make rare-earth ion-doped thin-film lithium niobate an ideal material to implement on-chip optical filters utilizing photonic cavities.

In this work, we demonstrate cavity-enhanced spectral filtering based on rare-earth ion-doped thin-film lithium niobate. We incorporate thulium ($Tm^{3+}$) ions into high-quality factor X-cut ring resonators patterned in thin-film lithium niobate. This cut is chosen as it is the same as that primarily used for thin-film lithium niobate modulators[43]. Due to the low propagation loss in the cavity, we attain a ring resonator structure with loss dominated by rare-earth ion absorption. By spectral hole burning at 4 K at the resonance frequency of a critically coupled mode of the resonator, in which all light is absorbed by thulium ions, we attain narrowband filters with linewidths ranging from 7 MHz with 13.0 dB of extinction, to 24 MHz with 20.4 dB of extinction. The 24 MHz-linewidth filter has an extinction ratio 70 times greater than that of a conventional waveguide of identical length. By reducing the temperature to 100 mK, we demonstrate an even narrower linewidth of 681 kHz, comparable to the narrowest filter bandwidth achieved in an integrated photonic device to-date[44], which is enabled by a mm-scale resonator in a low confinement silicon nitride platform at telecom wavelength. Our device features a footprint that is three orders of magnitude smaller and operates at visible wavelength where narrow bandwidth is more challenging due to higher propagation loss. Moreover, cavity coupling also enables creation of filter functions that are unattainable using bulk crystals or waveguides. For example, using an under-coupled cavity we show a bandstop filter as opposed to a bandpass filter. Such narrowband, high extinction ratio integrated spectral filters could serve as fundamental components for optical signal processing and optical memories on-a-chip.

**Physical System and Device Design**
Thulium ions doped in lithium niobate exhibit a quasi-three level energy structure, as shown in Fig. 1A. This configuration includes a ground level $^3H_6$, an excited level $^3H_4$, and an intermediate level $^3F_4$. The transition between the lowest energy level in $^3H_4$ multiplet and the ground level $^3H_6$ features a resonance wavelength of ~794 nm. Known for its strong absorption[45] and long coherence time of 1.4 $\mu s$ at a temperature of 4K[31], this transition is extensively studied for spectral hole burning and quantum information applications. In this work, we utilize the 794 nm optical transition inhomogeneously broadened by an ensemble of thulium ions to burn spectral holes by optically pumping the ground level population to the $^3H_4$ level. Such spectral holes have biexponential decay process determined by a ~70%-weighted $^3H_4$ level lifetime ($162 \pm 16 \, \mu s$) and ~30%-weighted $^3F_4$ level lifetime ($3 \pm 1.6 \, ms$)[46].

Fig. 1B shows a schematic of the proposed design of the cavity-enhanced spectral filter based on a ring resonator. The ring resonator consists of a closely spaced ring and a bus waveguide patterned in thin-film lithium niobate doped with $Tm^{3+}$ ions at a density of 0.1%. Light is coupled in and out of the resonator through grating couplers at both ends of the bus waveguide. Part of this light is coupled to the ring, as quantified by the coupling rate $\kappa_e$, which is determined by the gap between the bus and ring waveguides. As the light propagates around the ring resonator, the ring loss rate $\kappa_{in} = \kappa_{ion} + \kappa_p$ describes how fast the light is attenuated due to the dominant $Tm^{3+}$ ion absorption ($\kappa_{ion}$) and an additional low propagation loss ($\kappa_p$), where $\kappa_{ion} \gg \kappa_p$.

A critical coupling condition is achieved when $\kappa_e = \kappa_{in}$, in which all the light is coupled from the bus waveguide to the ring and lost during propagation, resulting in zero transmission to the output grating coupler. We achieve a cavity enhanced bandpass filter by utilizing a resonance mode (at ~794 nm wavelength) that is initially critically coupled and burning a spectral hole at the center of the mode so that $\kappa_{ion} = 0$ and $\kappa_e \gg \kappa_{in}$, resulting in high transmission at the burning wavelength. Meanwhile, the remaining incident light displays zero transmission, leading to a high extinction ratio for the bandpass filter. In contrast, for a conventional waveguide of the same length, absorption by the thulium ions is limited by the length of the device, which leads to a low extinction ratio (Fig. 1C). Although increasing the waveguide length or the ion doping concentration could enhance the ion absorption and improve the extinction ratio, a waveguide with high optical depth could experience higher power broadening in the front than the back. Moreover, a longer waveguide requires a larger device footprint, which is unfavorable for on-chip integration.

We first design the geometry of the thin-film lithium niobate ring and bus waveguides to maintain single transverse mode light operation with minimal propagation loss. For both waveguides, we use a rib design consisting of a strip on a slab structure, as shown in Fig. 1D. We choose the height of the strip and slab to be 140 and 160 nm, respectively. The sidewall angle is assumed to be 62° with respect to the horizontal direction, based on a previously reported fabrication approach[41,43]. A narrow width waveguide is beneficial for single mode light operation (i.e., the fundamental transverse electric (TE) mode) and strong interaction with ions. However, a larger width helps decrease the propagation loss by reducing the overlap between the optical mode and the waveguide sidewall. As propagation loss in the bus waveguide does not affect the total loss rate ($\kappa_{in}$) in the ring, we set the width to be 450 nm, with a bending radius of 40 $\mu m$, which supports only the fundamental TE mode at the target wavelength of 794 nm, according to finite difference time domain (FDTD) simulations[47]. However, for the ring waveguide, both single mode operation and minimal propagation loss are desired. Considering this trade-off, we use a ring design that tapers from a 650 nm wide waveguide down to a narrower width of 450 nm in the coupling region, which we set to be 30 $\mu m$ long. FDTD simulations of this coupling region verify that single mode coupling occurs at 794 nm. We also choose a large bending radius of 100 $\mu m$ for the ring waveguide to minimize the bending loss, which is $4 \times 10^{-6}$ dB/m according to simulations. Based on these parameters, the lengths of the bus and ring waveguides are 188 $\mu m$ and 1000 $\mu m$, respectively.

We then determine the coupling gap between the bus and ring waveguides to achieve the desired initial critical coupling condition ($\kappa_e = \kappa_{in}$). In addition to the length of the ring (1000 $\mu m$), $\kappa_{in}$ is determined by the configuration of the lithium niobate chip. We choose an X-cut configuration as it is widely used for modulators[43]. This configuration features an anisotropic absorption with respect to the optical axis (c-axis), where the ion polarization with $E \perp$ c-axis (waveguide ‖c-axis) exhibits strong absorption (3.5 dB/mm) at 794 nm, while its orthogonal direction $E \parallel$ c-axis (waveguide $\perp$ c-axis) shows negligible absorption coefficient[48]. As a result, we achieve low-loss coupling by aligning the straight coupling regions of the waveguides perpendicular to the c-axis[46,48,49], enabling simpler operation and analysis (see Fig. 1B). Based on this alignment, we estimate $\kappa_{in}$ to be 51.1 GHz by calculating the ion absorption at 794 nm in one round trip through the ring waveguide, which is given by the product of the ion absorption coefficient (3.5 dB/mm)[46] and the length of the ring that overlaps with the c-axis (500 $\mu m$). Then, to determine the coupling gap value that enables $\kappa_e = \kappa_{in}$, we simulate the coupling rate ($\kappa_e$) of the 30 $\mu m$ long coupling region of the bus and ring waveguides as a function of the gap distance. Our simulations show a gap of 510 nm between the bus and ring waveguides should result in the critical coupling condition (see Supplemental Information Section 1).

We then fabricate the proposed device design using electron beam lithography and dry etching. The chip is composed of a 300 nm thin film lithium niobate wafer doped with 0.1% $Tm^{3+}$ ions over a 2-$\mu m$-thick silicon dioxide layer grown on an undoped bulk silicon substrate (see Methods). Using electron beam lithography, we patterned the proposed ring resonator design into the lithium niobate chip, and then removed 140 nm of the unpatterned region using inductively coupled reactive ion etching with Ar$^+$ plasma (ICP-RIE) (see Methods). Given our simulations, we fabricated multiple ring resonators with coupling gaps that ranged from 410 nm to 610 nm at 5 nm steps. This wide range allows us to investigate the critical coupling condition within fabrication intolerance, as well as explore cavity-enhanced filters that feature other initial coupling conditions (e.g., $\kappa_e < \kappa_{in}$, $\kappa_e > \kappa_{in}$). Fig. 1e shows an optical microscopy image of a representative ring resonator and scanning electron microscopy images of its coupling region and grating coupler. For comparison, we also fabricated conventional waveguides of 650 nm in width and 1000 $\mu m$ in length on the same lithium niobate chip.

**Results and Discussion**

The propagation loss of both the fabricated ring and conventional waveguide is estimated to be around 0.7 dB/cm at 794 nm. This value is obtained from a nearly critically coupled bare ring resonator with a much shorter coupling region but identical ring waveguide design, fabricated using the same methods on a separate undoped thin-film lithium niobate chip. The low propagation loss results in a high-quality factor of the bare ring resonator, measured to be $Q_L = 5.42 \times 10^5$, corresponding to an intrinsic quality factor of $Q_I = 1.1 \times 10^6$ at 794 nm (see Supplemental Information Section 2).

We first characterize the ring resonance modes using photoluminescence. When excited to the fifth level of the $^3H_4$ excited level multiplet[48] by a continuous wave (CW) laser at ~773 nm, the thulium ions decay to the $^3H_6$ level and photoluminescence at 794 nm (Fig. 1A). Using a ring resonator with a coupling gap of 550 nm, we measure the photoluminescence spectra of the $Tm^{3+}$ ions at 4 K at different excitation wavelengths between 773 and 774 nm. Figure 2A shows the photoluminescence emission of the device at excitations of 773.39 nm and 773.53 nm. For 773.39 nm excitation, narrow-linewidth photoluminescence peaks appear in the spectrum, which we attribute to the resonances of the ring resonator. Background photoluminescence emission appears in the spectrum taken at 773.53 nm excitation, which is off-resonant to the cavity. The resonance peaks are equally distributed with a free spectral range (FSR) of 0.263 nm. This value is consistent with the predicted free spectral range for the fundamental TE mode, given by $FSR = \lambda^2/n_g L = 0.268\ nm$, where $\lambda$ and $L$ represent the wavelength and total length of the ring waveguide, respectively, and $n_g$ is the simulated TE mode group index ($n_g = 2.33$). We therefore attribute the observed resonance peaks to the fundamental TE modes of the ring waveguide.

We next measure the transmission spectrum of the device. To avoid affecting the ion absorption by hole burning, we use a weak superluminescence diode (SLD) source (10 nW), with a center wavelength of 810 nm and a bandwidth of 30 nm. Figure 2B shows the measured transmission spectrum for the same device used in panel A. The device exhibits a series of spectral dips whose free spectral range (0.263 nm) is consistent with that of the photoluminescence spectrum. Within the wavelength range of the thulium ion absorption (orange line in Fig. 2B), we observe deep spectral dips. The transmission spectrum of the dips at wavelengths that correspond to the maximum absorption of the ions drop to nearly zero, suggesting that we are close to critical coupling ($\kappa_e = \kappa_{in}$). The contrast of the dips outside the absorption range of the thulium ions is very small, indicating that these modes are over-coupled ($\kappa_e \gg \kappa_{in} = \kappa_p$).

To further confirm whether the critical coupling condition is met and determine the exact values of $\kappa_e$ and $\kappa_{in}$, we conduct a higher resolution transmission measurement using a narrowband tunable laser. We focus on the resonance mode at 794.535 nm, as indicated by the blue arrow in Fig. 2B. We scan the wavelength of an attenuated tunable CW laser (resolution 0.001 nm) and measure the power of the transmitted light as a function of the wavelength, as shown in Fig. 2C. We fit the resulting transmission spectrum of the 794.535 nm resonance mode to the all-pass ring resonator transmission equation[50] (blue line, Fig. 2C):

$$T = \frac{r^2 + a^2 - 2ar\cos\varphi}{1 + r^2a^2 - 2ar\cos\varphi} \tag{1}$$

where $r$ is the amplitude coefficient of light that remains propagating along the bus waveguide at the coupling region, $a$ is the amplitude attenuation factor in the ring waveguide, and $\varphi$ is the round-trip phase given by $2\pi L/\lambda$. The transmission at the fitted resonance dip is T = 1% of the light coupled into the bus waveguide, which is close to zero transmission for the ideal critical coupling condition. From the fit, we determine the fitting parameters to be $r = 0.83$ and $a = 0.80$. Using these values, we calculate $\kappa_e/2\pi$ and $\kappa_{in}/2\pi$ as 7.43 GHz and 8.7 GHz via their relations[51]:

$$\kappa_e = \frac{2c \cdot \cos^{-1}\left(\frac{2r}{1+r^2}\right)}{n_g L} \tag{1}$$

$$\kappa_{in} = \frac{2c \cdot \cos^{-1}\left(\frac{2a}{1+a^2}\right)}{n_g L} \tag{2}$$

where $c$ is the vacuum light speed and $n_g$ is the group index of the TE mode. The extracted ring loss rate is also comparable to the predicted $\kappa_{in}$ value (52.9 GHz) that we estimate from the measured propagation loss (0.7 dB/cm) and ion absorption (3.5 dB/mm at 794.535 nm, see orange line in Fig. 2B). Based on these results, we conclude that a near critical coupling condition is successfully achieved for this resonance mode at 794.535 nm in a ring resonator with a coupling gap of 550 nm, which is close to the value predicted by simulations.

To create a cavity-enhanced bandpass filter, we burn a deep spectral hole in the critically coupled resonance mode at 794.535 nm. We generate a pulse of 50 $\mu s$ duration to excite the ions from the $^3H_6$ ground level to the $^3H_4$ excited level, followed by two weak probe pulses to investigate the transmission spectrum of the resulting spectral hole (see Methods). To maximize the extinction ratio of the filter, we use a burning power of 250 nW to remove all ion absorption at the target wavelength and burn the deepest possible hole. The frequency of the probe light is swept through a 100 MHz bandwidth around the burning frequency and the transmitted light is measured as a function of frequency to obtain the transmission spectrum of the filter.

Fig. 3A shows the measured relative transmission spectrum of the device (blue line), which we define as the transmission spectrum $T(\omega)$ normalized to the minimum background transmission $T_{back}$ measured without any burned spectral hole. This spectrum features a strong transmission peak at the burning wavelength, with the corresponding extinction ratio given by the maximum value at the peak. By fitting the data with a Lorentzian function, we extract a narrow linewidth of $24 \pm 0.9$ MHz and a high extinction ratio of $110 \pm 1.1$ (20.4 dB). These results demonstrate that after spectral hole burning the ring resonator functions as a narrow bandpass filter for the targeted wavelength.

For comparison, we also create a bandpass filter by spectral hole burning a conventional lithium niobate waveguide of the same length as the ring resonator (1000 $\mu m$). To remove all ion absorption at the target wavelength, we use a burning power that is 10-times as strong as the power used in the ring resonator (2500 nW) to compensate for the ~10-times power enhancement in the ring waveguide due to resonance[52] (see Supplemental Information Section 3). The orange line in Fig. 3A shows the relative transmission spectrum of the waveguide, which features a linewidth of $18 \pm 0.7$ MHz and a low extinction ratio of $1.52 \pm 0.01$. The theoretical maximum extinction ratio is determined by the optical depth (OD) of the device, which we can calculate from the product of the ion absorption coefficient (Fig. 2B) and the length of overlap of the waveguide with the c-axis of the lithium niobate chip (500 $\mu m$). From these values, we find OD = 0.43, which gives a theoretical maximum extinction ratio of $e^{OD} = e^{0.43} = 1.54$. Therefore, the measured extinction ratio (1.52) of the conventional waveguide is already near the maximum and cannot be further improved, even using a higher burning power. In comparison, the cavity-enhanced bandpass filter features a 70-times higher extinction ratio while maintaining the same device footprint.

We also investigate the dependence of the extinction ratio of the cavity-enhanced filter on the hole burning power. We use the same conditions for the spectral hole burning experiment, fixing the burning pulse duration to be 50 $\mu s$ while changing the burning power. Figure 3B shows the measured power-dependent extinction ratio of spectral holes burned in the same critically coupled resonance mode in the 550 nm gap ring resonator (black), which we compared to a 2000 $\mu m$ conventional waveguide (blue). The extinction ratio of the spectral hole increases with the burning power in both cases, as more ions are excited at a higher power until the deepest hole is created. For the cavity-enhanced filter, the hole extinction ratio grows from 20 to 110 as the power increases from 30 nW to 250 nW, while for the case of the conventional waveguide, the extinction ratio increases from 2.3 and gradually saturates at 3.2 for a power of 100 nW to 2300 nW. These results show that the critically coupled resonance mode greatly enhances the extinction ratio of the cavity-enhanced filter. Even at the smallest investigated burning power (30 nW), the cavity-enhanced filter features a 6-times higher extinction ratio compared to the maximum of the conventional waveguide. We would expect to see a saturation trend for the cavity-enhanced filter at higher burning powers. However, in our current setup, the resonance modes in the ring resonator exhibit thermo-optic shifts at high power[53]. As a result, we were not able to increase the burning power over 250 nW without losing the initial critical-coupling condition. Nevertheless, it is worth noting that a higher power and further improvement on extinction ratio is possible by compensating this red shift using DC tuning[54].

We also explore the power-dependent linewidths of both the cavity-enhanced filter and waveguide filter. Figure 3C shows the linewidths of the same spectral holes investigated in Fig. 3B, at different burning powers in both the ring resonator (black) and a 2000 $\mu m$ waveguide (blue). In Fig. 3C, the hole linewidth increases at higher burning powers for both the ring resonator and waveguide, which can be explained by the power-broadening effect of the spectral hole[55,56]. The linewidth of the cavity-enhanced filter increases from 7 MHz to 24 MHz as the power increases from 30 nW to 250 nW, while the linewidth of the conventional waveguide filter grows from 5 MHz to 17 MHz at powers from 100 nW to 2300 nW, which can be fitted to the broadening model described in [55] (see Supplemental Information Section 4). The minimum achievable linewidth is extracted to be 2.3 MHz from the zero-excitation intercept of the fitted curve, which is limited by both the homogeneous linewidth and the laser linewidth (see Supplemental Information Section 4). For the same burning power, the cavity enhanced filter features a linewidth that is approximately three times as broad as that of the conventional waveguide, which we attribute to the enhanced burning power via the ring resonance. Thus, although the filter bandwidth of the cavity is slightly degraded relative to a waveguide, the linewidth broadening is small compared to the significant improvement in the extinction ratio. By utilizing different burning powers, this tradeoff between the filter extinction ratio and linewidth can be managed for different applications.

The linewidth of the cavity-enhanced filter can be further reduced by going to lower temperature to reduce phonon broadening[55]. Figure 4A shows a cavity enhanced filter linewidth attained at 100 mK using a 30 nW burning power. In contrast, Fig. 4B shows one of the 4 K cavity-enhanced filters previously demonstrated in Fig. 3, burned with the same 30 nW power. By fitting the filter transmission spectra to a Lorentzian function, we extract the filter linewidth to be 681 kHz, which is an order of magnitude improvement over the 7 MHz linewidth observed at 4 K. Notably, this linewidth is among the narrowest filters demonstrated in an integrated photonic device[44,57,58]. Moreover, the filter bandwidth is not a fundamental limit at 100 mK, it is determined by the linewidth of the burning laser. Figure 4C shows a photon echo measurement (see Methods), which yields a coherence time of $11.3 \pm 0.28 \,\mu s$. This coherence time corresponds to a fundamental minimum filter linewidth of 56 kHz (see Supplemental Information Section 4). The coherence time could be improved even further by applying a magnetic field, which reduces this fundamental limit to 5.4 kHz[59].

Apart from the frequency filtration effect, a bandpass filter also imparts a group delay on an incident light pulse due to dispersion[60]. To demonstrate this time domain response of the cavity-enhanced bandpass filter, we first burn a spectral hole in the critically coupled resonance mode and then send a weak Gaussian-shaped probe signal of duration 1 $\mu s$ with the same carrier frequency through the hole (see Methods). For the spectral hole burning, we use a burning power of 150 nW, which gives a good balance in the resulting extinction ratio (70) and hole linewidth (16 MHz). Figure 5A shows the transmission spectrum of the burned spectral hole, and its absorption spectrum, calculated by $\alpha(\omega) = -\ln[T(\omega)]$ (inset). The spectral hole features a linewidth of $\Delta = 16 \, MHz$ and a hole depth of $d = 4.25$.

Figure 4B shows the measured arrival time of the probe pulse transmitted through the burned spectral hole. For comparison, we also measure the arrival time of a reference probe pulse transmitted through a very broad spectral hole (>1 GHz) burned with 80 $\mu W$ power, which is considered to have negligible group delay (orange, see Methods). By fitting both pulses with Gaussian functions, we extract the delay of the probe pulse center to be $46 \pm 7$ ns, which is consistent with the theoretical delay predicted from the hole depth $d$ and linewidth $\Delta$ as: $\Delta t = d/2\pi\Delta = 42\ ns$ [37]. This consistency between the measured time response of the cavity-enhanced filter with the predicted time delay further validates its properties as a bandpass filter.

A distinctive advantage of using cavity-coupled emitters is the ability to produce reconfigurable filtering behavior by controlling the cavity coupling rate. As an example, in contrast to the bandpass filter, we demonstrate a method to generate a bandstop filter (Fig. 6A). To achieve this type of filter, we employ a resonance mode that is initially in an under-coupled condition ($\kappa_e < \kappa_{in}$) with $r > a$ according to Eq. (2-3). Under this condition, the resonance wavelength exhibits a large transmission, as described by substituting $\varphi = 0$ in Eq. (1):

$$T = \frac{(r-a)^2}{(1-ra)^2} \qquad (3)$$

By burning a spectral hole in the resonance mode, we could decrease $\kappa_{ion}$, reaching the critical coupling condition $\kappa_e \approx \kappa_{in} = \kappa_{ion} + \kappa_p$ at the burning wavelength, resulting in $r = a$ and zero transmission in Eq. (4). This reduced transmission at the burning wavelength enables a bandstop filter.

To demonstrate this concept, we investigate another ring resonator that features a larger coupling gap of 605 nm, resulting in an under-coupled resonance mode at 794.13 nm with $\kappa_e/2\pi$ = 2.99 GHz and $\kappa_{in}/2\pi$ = 10.31 GHz (see Supplemental Information Section 5). We burn a spectral hole at the center wavelength of this resonance mode using a burning power of 100 nW. Figure 5C shows the transmission and absorption (inset) spectra of the device after the hole burning, which features a transmission dip at the center frequency, allowing it to function as a bandstop filter. By fitting the spectrum to a Lorentzian function, we extract the linewidth to be 4.2 MHz and the extinction ratio to be 4.4, with a background transmission of 30%. This smaller extinction ratio compared to the cavity-enhanced bandpass filter is mainly attributed to the challenges in optimizing the spectral hole burning process. While increasing burning power can maximize the extinction ratio for a bandpass filter, achieving optimal condition $\kappa_e = \kappa_{in}$ for a bandstop filter demands precise control over the burning power to remove a specific amount of ion absorption. However, there is no fundamental limit on the maximum achievable extinction ratio for this bandstop filter, as the optimal condition $\kappa_e = \kappa_{in}$ gives zero transmission at the burning wavelength, resulting in an infinitely large extinction ratio. Alternatively, we can utilize an even more under-coupled resonance mode that has a larger difference between $\kappa_e$ and $\kappa_{in}$ ($\kappa_e \ll \kappa_{in}$) to improve the transmission of the un-burned wavelengths, which would also further increase the extinction ratio.

The response time of the bandstop filter features interesting behavior. As a narrow resonance dip in the under-coupling regime, the bandstop filter induces a negative slope of phase shift, which gives rise to a negative group delay[61]. Therefore, unlike the bandpass filter which delays the input pulse, the bandstop filter advances the pulse in time[60]. Similar to faster than light group velocities, this behavior may seem counter-intuitive because it suggests that the pulse leaves the cavity before it enters[62]. It should be noted, however, that this effect does not violate causality[63].

To demonstrate negative group delay, we send a weak (5 nW) Gaussian-shaped probe pulse through the filter after a 10 $\mu s$ delay. Figure 5D shows the measured arrival time of the probe pulse transmitted through the bandstop filter (blue), as well as a reference probe pulse (orange). By fitting both pulses with Gaussian functions, we extract the advance of the pulse of the bandstop filter to be 55 ± 7 ns. This measured result is also consistent with the predicted advance time calculated from the hole depth $d = 1.42$ and linewidth $\Delta = 4.2 MHz$ as $\Delta t = d/2\pi\Delta = 53\ ns$.

Besides demonstrating a bandstop filter at the center of the resonance mode, we also explore the filter behavior when the burning wavelength is detuned from the resonant wavelength. To achieve this, we utilize a CW laser detuned from the resonance wavelength (794.13 nm) by -0.006 nm (794.124 nm) and 0.006nm (794.136 nm) (Fig. 5B, red arrows). The transmission spectra of the resulting detuned bandstop filters are shown by the blue lines in Fig. 5E and F, respectively. In contrast to the Lorentzian shape exhibited when the filter aligns with the cavity resonance, the detuned bandstop filters feature Fano-shaped transmission profiles. The detuned filters are symmetric because their burning wavelengths are blue and red-shifted by the same amount with respect to the resonance wavelength. This Fano-shaped distortion is attributed to a dispersion effect originating from the narrow spectral hole burned in the atomic absorption. This dispersion effect induces a frequency-dependent change in the real refractive index component, which subsequently alters the original transmission spectrum[64,65]. By calculating the resulting real refractive index component using the Kramers-Kronig relations based on the absorption spectrum of the ion absorption, we model these Fano-shaped transmissions, as shown in the black line in Fig. 5E, F (see Supplemental Information Section 6), which are in good agreement with the measured data. This model is instrumental for the future implementation of more complicated cavity-enhanced reconfigurable filtering functions, which require a series of bandpass or bandstop filters within a resonance mode[66].

The lifetime of both the bandpass and bandstop filters can be improved by modifying the hole burning procedure. By extending the burning pulse to 5 ms, much longer than the $^3H_4$ lifetime, we can drive most of the population to $^3F_4$ level, which persists for ~3 to 7 ms[46,48]. With this burn pulse sequence, the cavity-enhanced filter exhibits a single exponential decay with a time constant of 4.5 ms (see Supplemental Information Section 7). We note that this exponential decay can be further mitigated by using a backwards pumping scheme, where a constant burn pulse is sent in the opposite direction to the signal to sustain a persistent filter. This configuration enables continuous wave operation while still allowing complete separation between the pump and signal.

Another important consideration is the noise created by the burn pulse due to residual spontaneous emission from $^3H_4$. This noise decays on the time scale of $^3H_4$ lifetime, much shorter than the 4.5 ms filter lifetime achieved with 5 ms burning pulse. To investigate if the noise level would allow single-photon filtering, we performed a noise analysis (see Supplemental Information Section 8). By waiting 380 $\mu s$ after the burn pulse, the spontaneous emission created by the burn pulse drops well below a single photon per filter lifetime, while the filter only decays by ~10 %, indicating that the filter is suitable for quantum applications.

This noise level can be further significantly improved when we apply a magnetic field and utilize the long lived hyperfine state for hole burning, which features a lifetime of tens of minutes[59]. By performing the noise analysis using a magnetic field of 150 G along the c-axis, we observe no measurable decay of the filter extinction ratio over a 10 ms timescale, while the noise signal decays to unmeasurable levels after 1 ms. Therefore, under this condition, we can burn a persistent hole with virtually no noise, enabling quantum operation under a broad range of operating conditions.

While the demonstrated filters already feature favorable properties, there is still room for further improvements. The extinction ratio of the cavity enhanced filter is currently limited by the precision that we can achieve critical coupling. By precisely controlling the burning pulse in an under-coupled device, we could tune the loss to achieve a more accurate critical coupling regime and improve the extinction ratio. Alternatively, we could leverage lithium niobate's strong electro-optic coefficient to actively tune the coupling condition, either by tuning the resonance frequency[54] or the coupling strength[67]. These approaches could be used to compensate for fabrication imperfections, such as variations in etch depth or coupling gaps, and enable a precise critical coupling condition with high device yield.

The maximum transmission of the cavity-enhanced filter shown in Fig. 3A is approximately 56%, which means we have a loss of 2.4 dB. This loss is resulted from a nonzero ring loss $k_{in}$ after the hole burning, which is not fundamental and could be significantly improved. Our calculation attributes 1.6 dB loss to residual ion absorption ($k_{ion} = 4.46\ GHz$) because of imperfect hole burning, while the remaining 0.8 dB is attributed to waveguide propagation loss ($k_p = 2.1\ GHz$) (see Supplemental Information Section 8). The residual ion absorption could be reduced by increasing the burning power to create a deeper spectral hole. Currently, due to the lack of active modulations to compensate for resonance red shift, the maximum burning power is limited to only 250 nW. Alternatively, we could extend the spectral hole lifetime by applying a magnetic field[45], thus reducing the population decaying back to ground state that contributes to the residual absorption. The propagation loss could be further reduced as well by depositing a SiO$_2$ cladding layer, utilizing a wider ring waveguide geometry, or leveraging better fabrication techniques[69].

Lastly, the allowed input signal bandwidth (e.g., the stopband of the bandpass filter) is determined by the linewidth of the specific resonance mode utilized for hole burning, which is ~16 GHz and ~13 GHz for the demonstrated bandpass and bandstop filters, respectively. While these bandwidths may be insufficient for certain applications, they are practically useful for applications requiring MHz or kHz resolution[18,20,70]. Moreover, this bandwidth can be extended by cascading another coarse passive bandpass filter before our device to create a two-step filter. Depending on the applications, we can also adjust the current device into a larger ring design to increase the operation bandwidth.

In conclusion, we demonstrate cavity-enhanced narrowband spectral filters with high extinction ratio using rare-earth-ions doped in thin-film lithium niobate. We achieve a filter bandwidth as narrow as 681 kHz, comparable to the narrowest filter bandwidth demonstrated in an integrated photonic device[44], in a device footprint that is three orders of magnitude smaller. By controlling the cavity coupling rate, we also show reconfigurable filtering through the creation of a bandstop filter. Our device operates at a wavelength $< 1\ \mu m$, where achieving narrowband integrated filters is especially challenging due to higher propagation loss. Our approach also circumvents the undesirable mode splittings[71–73] commonly experienced by high-Q integrated photonic cavities. Another important advantage of our filters is the ability to tailor the spectral hole burning process to programmably create arbitrary-shaped filters, such as spectral gratings. Our demonstration of the fundamental element – a single spectral filter- is the first step to realizing these more complex filter functions in this versatile integrated platform. The co-integration of these filters with electro-optic devices, such as frequency shifters[43,74], could further broaden the utility of rare-earth spectral tailoring[75].

## Methods

### Device Fabrication:

We fabricate the devices using 300 nm thick X-cut LN thin films doped with 0.1% $Tm^{3+}$ on 2 $\mu m$ of silicon dioxide on silicon substrates. To obtain the doped thin film wafers, $Tm^{3+}$ ions are first volume-doped into a bulk lithium niobate crystal boule through Czochralski growing method ((OST photonics)[78]. A thin-film layer is then fabricated from the bulk crystal using a "Smart-Cut" technology. In this process, a cleavage plane is first defined at the desired 300 nm thickness via high-dose implantation of $He^+$ (or $H^+$) ions, bonded to the insulating silicon substrate with an oxide layer, and finally annealed to split the crystal along the defined cleavage plane. Next, to fabricate the devices, we use standard electron beam lithography to define patterns in hydrogen silsesquioxane (HSQ) resist with multipass exposure. The patterns are subsequently transferred into the LN thin-film using a commercial inductively coupled plasma reactive ion etching (ICP-RIE) tool. We use $Ar^+$ plasma to physically etch LN where the plasma power and chamber condition are tuned to minimize surface redeposition of removed LN and other contaminations present in the chamber. After etching, we remove redeposition using RCA SC-1 cleaning procedure. Finally, we anneal the sample at atmospheric pressures in $O_2$ at 520 ℃ for two hours, to improve the crystallinity of thin-film lithium niobate.

### Spectral hole burning measurement:

For the spectral hole burning measurements, we generate one burn pulse and two weak probe pulses by modulating a narrowband continuous wave laser (Msquared Solstis) utilizing two fiber-coupled acousto-optic modulators (AOM, Brimrose Corp), which are controlled by an arbitrary waveform generator (Keysight). These two AOMs provide independent control over the power, duration and frequency of the burn and probe pulses. We set the burn pulse duration at a fixed 50 $\mu s$ and vary its power from 50 nW to 2500 nW for different measurements, while maintaining the probe pulses at a constant 3 nW to avoid affecting the hole population. To probe the transmission of the spectral hole, we slowly sweep the frequency of the probe pulses from -50 MHz to 50 MHz with respect to the burning frequency. The first probe pulse is shortly delayed from the burn pulse by 10 $\mu s$, to minimize the spectral hole recovering from excited level population decay. Then we delay another 10 *ms* to send the second probe pulse, which is well exceeding the $^3H_4$ and $^3F_4$ lifetimes to ensure complete reset of the burned spectral hole to the original ground level population. We record the probe pulses on a single photon counting module (Excelitas Technologies Inc). The normalization of the first probe pulse to the second allows for the measurement of the transmission spectrum of the spectral hole.

**Two-pulses photon echo measurements:**
We utilize two laser pulses with 13 $\mu W$ power to coherently rotate the atomic population. We set the duration of first laser pulse to be 60 ns, which corresponds to a pulse area of $\pi/2$. As a result, it coherently rotates the population to a superposition of the ground and excited state. The second pulse, delayed by time $\tau$, has a duration of 100 ns and a corresponding pulse area of $\pi$ that rephases the atomic population. This rephasing results in a photon echo delayed by a time $\tau$ relative to the second pulse. By measuring the strength of the photon echo as a function of the delay time $\tau$, we can determine the optical coherence time.

**Slow and fast light measurements:**
For the slow and fast light measurement, we use a setup similar to the one used for spectral hole burning measurements, but with one burn pulse and one weak probe pulse. Instead of sweeping the probe frequency, we fix it to match the burn frequency and modulate the probe amplitude into a Gaussian shape. We set the full width at half maximum of the pulse to 1 $\mu s$, yielding a probe bandwidth (1/1 $\mu s$ = 1 MHz) much smaller than the created hole. This condition ensures that the probe pulse transmits through the spectral hole without distortion, allowing us to describe the shift in arrival time as $\Delta t = \pm d/2\pi\Delta$[37]. To measure the shift in probe pulse arrival time, we generate a reference probe pulse by repeating the same measurement with a much higher burning power of 80 $\mu W$, which leads to a very broad spectral hole (>1 GHz) causing negligible changes in the group velocity. The resulting reference probe pulse is therefore considered unshifted in arrival time and serves as a benchmark for evaluating the time delay or advance.

## Acknowledgements
The authors would like to acknowledge support from the National Science Foundation (Grant No. OMA2137723). The Waks group would also like to acknowledge financial support from the National Science Foundation (Grant No. OMA1936314), the U.S. Department of Defense contract (Grant No. H98230-19-D-003/008), and the Maryland-ARL Quantum Partnership (Grant No. W911NF1920181). The Lončar group would also like to acknowledge financial support from the National Science Foundation (Grant No. EEC1941583) and the Army Research Office MURI (Grant No. W911NF1810432).

## Data availability Statement
All data used and analyzed for this current study are available within the article and its Supplementary Information file, or are available from the authors upon request.

## Competing interests

All authors declare no financial or non-financial competing interests.

**Author contributions**
Y.Z. and N.S. conceived the experiment. D.R. fabricated the device. Y.Z. performed the experiment. D.F., Y.J. and S.D. supported in setting up the experiment. Y.Z. analyzed the data. Y.Z., E.W. and N.S. prepared the manuscript and all authors reviewed it and discussed the results. E.W. and M.L. supervised the experiment.

**Figure**

Figure 1

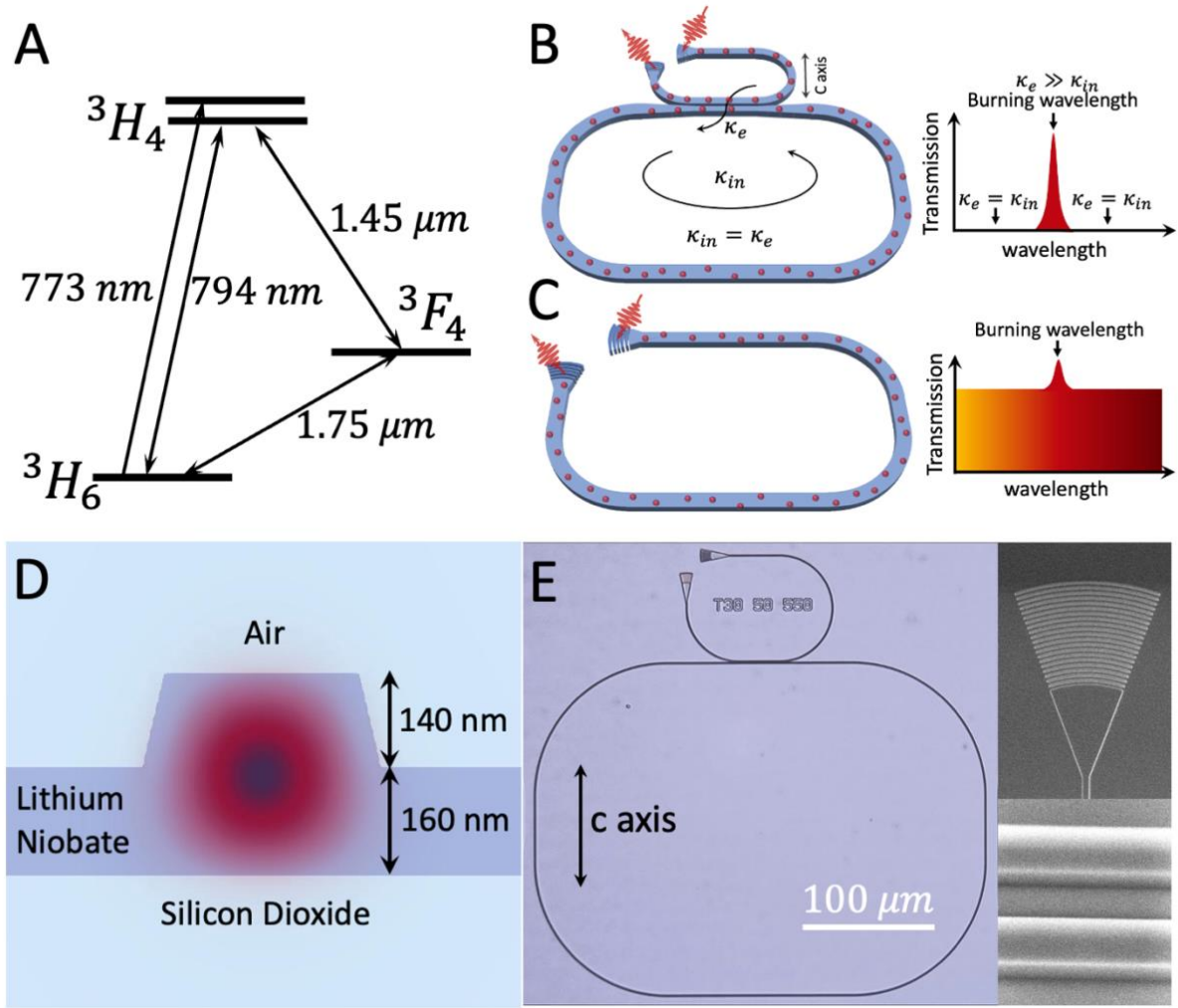

**Fig. 1** (A) A simplified energy level structure and approximate transition wavelengths of $Tm^{3+}$ ions in thin-film lithium niobate. (B) Schematic of the proposed cavity-enhanced spectral filter design based on a critically coupled ring resonator. By spectral hole burning at the target wavelength, the initial critically coupled condition can be changed into strongly over-coupled condition ($\kappa_e \gg \kappa_{in}=\kappa_p$) at the burning wavelength, which gives high transmission. Meanwhile, the light at the unburned wavelength remains zero transmission, which enables a high extinction ratio bandpass filter. (C) Schematic of a bandpass filter based on a conventional waveguide of the same length of the ring. Due to the small absorption limited by the length of the waveguide, the filter features a low extinction ratio. (D) Schematic of the rib waveguide design cross section and the simulated fundamental TE mode inside the waveguide. (E) An optical microscope image of a representative fabricated ring resonator and scanning electron microscopy images of its coupling region and grating coupler. The long coupling region allows reaching high coupling rates ~ 50 GHz at 794 nm wavelength.

Figure 2

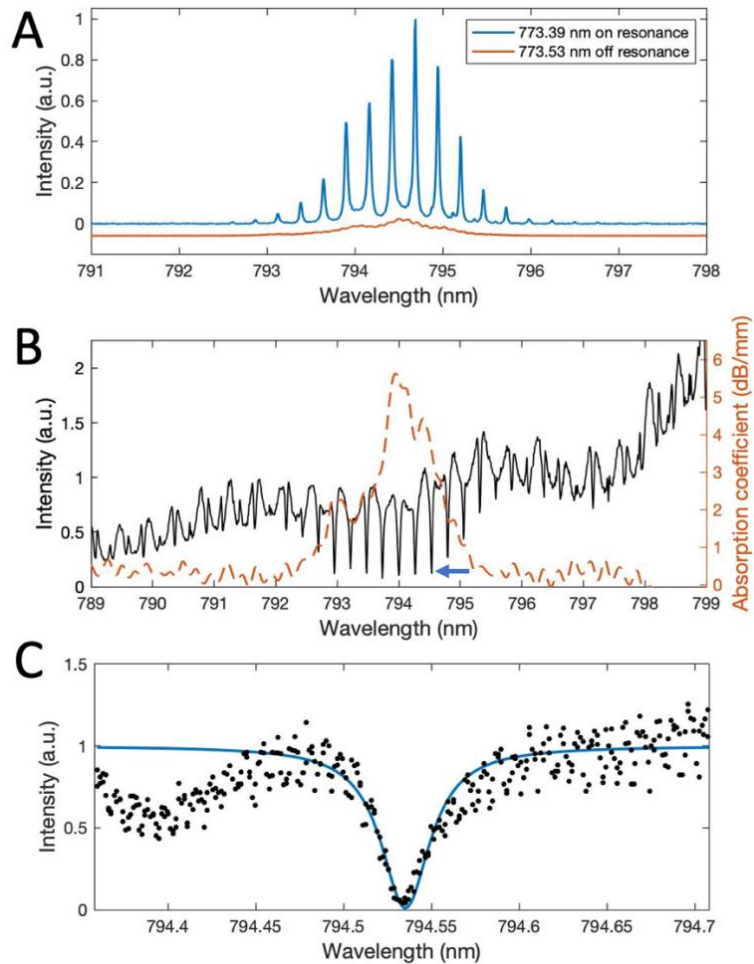

**Fig 2**. (A) The photoluminescence spectrum of a ring resonator with a 550 nm coupling gap, excited by light that is on-resonant (773.39 nm, blue) and off-resonant (773.53 nm) to the ring. The spectrum is normalized to the maximum photoluminescence intensity. (B) The transmission spectrum of the same ring resonator (black), normalized to the intensity of non-resonant wavelength at ~794 nm. The overall shape of the spectrum is due to the profile of the SLD source we used. The absorption spectrum of the $Tm^{3+}$ ions measured from a conventional waveguide (orange) is also shown for comparison. (C) The transmission spectrum of the resonance mode at 794.535 nm (blue arrow in (B)), fitted with a Lorentzian curve and normalized to the fitted constant background intensity. The small resonance at 794.4 nm is due to Fabry-Perot interference between optics in the measured setup. The background noise is due to Fabry-Perot interference between optics in the measurement setup.

Figure 3

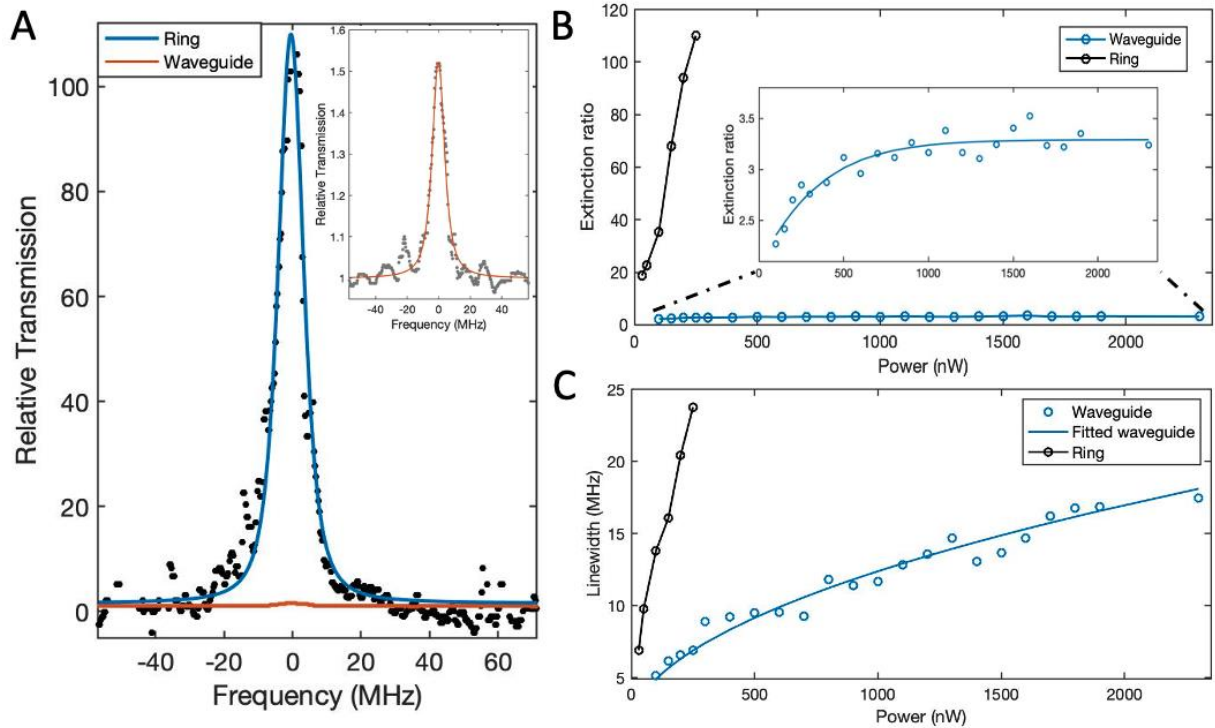

**Fig. 3** (A) The relative transmission spectra of a deep spectral hole burned with 250 nW burning power in the critically coupled resonance mode in a 550 nm gap ring resonator (blue), and a spectral hole burned with 2500 nW burning power in a 1000 $\mu m$ conventional waveguide (orange and inset). Both spectra are normalized to the minimum background transmitted power. The noises are likely due to thermally driven photodetection dark count. (B) The extinction ratio of the cavity-enhanced (black) and conventional waveguide filters (blue), as a function of the burning power. Note, for this measurement we use a longer conventional waveguide (2000 µm) to achieve a higher maximum extinction ratio compared to in Fig. 3A (orange) to better observe the trend of the extinction ratio as a function of the burning power. (C) The linewidth of the cavity-enhanced filter (black) and the conventional waveguide filter (blue, 2000 µm in length), as a function of the burning power. The fluctuations are likely due to the laser instability.

Figure 4

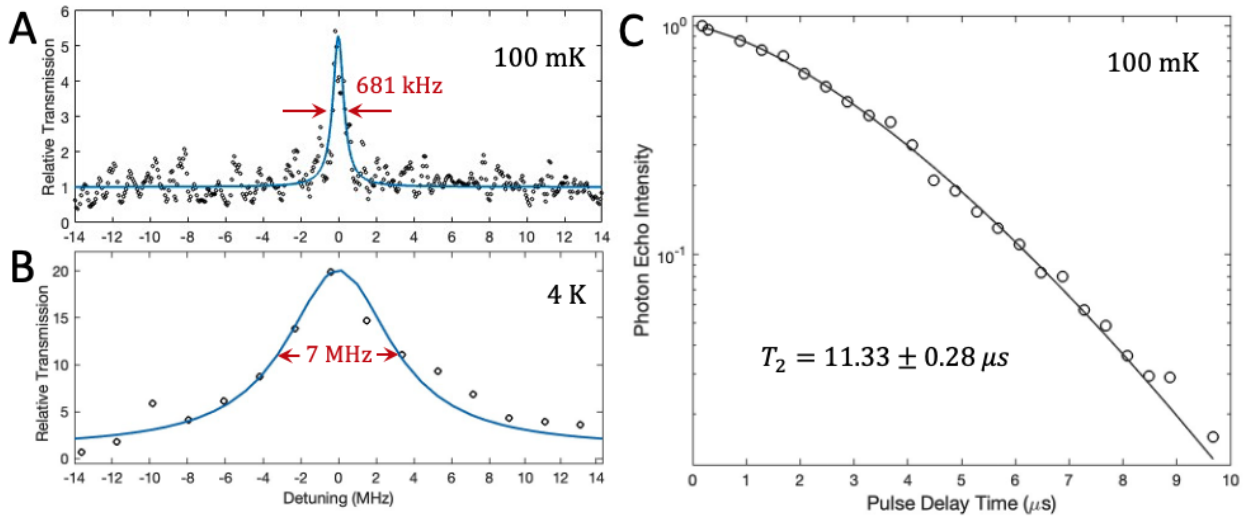

**Fig.4** (A,B) The relative transmission spectra of a cavity-enhanced filter created with 30 nW burning power at 100 mK and 4 K, respectively. Note that the extinction ratio of the filter at 100 mK is lower due to slightly degraded critical coupling condition after we reloaded the sample in the dilution fridge, which likely resulted from dirt or condensation, leading to increased propagation loss. (C) Two-pulse photon echo decay at 100 mK.

Figure 5

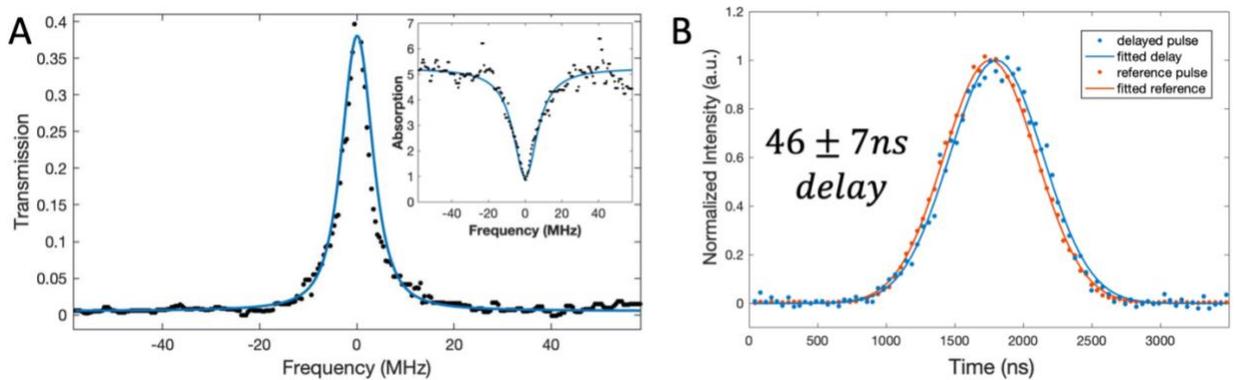

**Fig. 5** (A) The transmission and absorption (inset) spectra of a spectral hole burned with 150 nW burning power in a critically coupled resonance mode of the ring resonator. (B) The arrival time of a Gaussian-shaped weak probe pulse transmitted through the burned spectral hole shown in (A) (blue) is delayed by $46 \pm 7\,ns$, as compared to a reference probe pulse (orange).

Figure 6

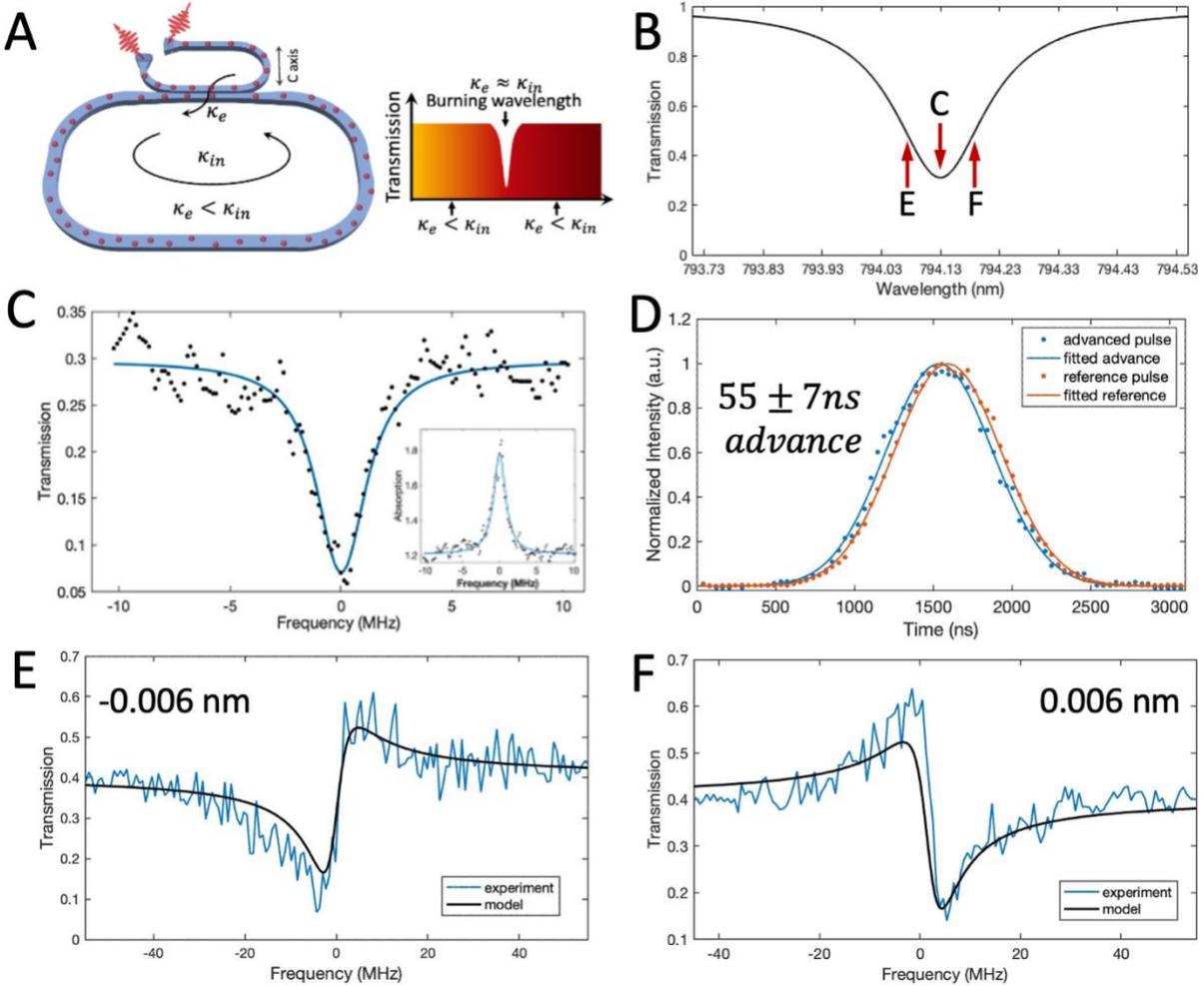

**Fig. 6** (A) A schematic of a bandstop filter created by burning a spectral hole in an initially under-coupled resonance mode. (B) The calculated transmission spectrum of an under-coupled resonance mode. (C) The transmission spectrum of a bandstop filter created by burning a spectral hole at the center of the resonance mode. (D) The arrival time of a probe pulse transmitted through the bandstop filter in (C) (blue), as well as an unshifted reference probe pulse (orange). (E, F) The transmission spectra of a Fano-shaped bandstop filter created by burning a spectral hole at a wavelength that is detuned from the resonance by (E) -0.006 nm and (F) 0.006 nm.

# Supplemental Information

**Section 1. Determination of the desired gap distance for the critical coupling condition**

We first estimate the ring loss rate based on the selected ring design geometry (1000 $\mu m$ circumference, 500 $\mu m$ overlap with c-axis). By calculating the ion absorption at ~794 nm in one round trip through the ring waveguide from the product of the ion absorption coefficient (3.5 dB/mm) and the absorbing length (500 $\mu m$), we obtain the amplitude attenuation factor of $a = 0.82$, and the ring loss rate of $\kappa_{in} = 51.0\ GHz$ (see Eq. (1)(3)).

To determine the desired coupling gap to achieve the critical coupling condition, we apply coupled mode theory[1] to analyze the coupling region of the ring resonator. In the straight coupling region, the evanescent mode coupling between the bus and the ring waveguide results in two new modes, a symmetric and an anti-symmetric mode, as shown in Fig. S1. The electric field amplitude splitting between the waveguides depends sinusoidally on the difference between the effective refractive indices of these modes, $\Delta n$, as: $E_{ring}/E_{bus} = sin(\frac{\pi \Delta n}{\lambda_0} L_c)$, where $L_c = 30\ \mu m$ is the coupling region length and $\lambda_0$ is the free space operating wavelength (794 nm). We simulate $\Delta n$ as a function of the top gap distance between the bus and ring waveguides, and calculate the resulting ratio between $E_{ring}$ and $E_{bus}$, as shown in Fig. S1 C. The critical coupling condition can be achieved when $E_{ring}/E_{bus} = \sqrt{1-a^2} = 0.57$ is satisfied, which corresponds to a gap distance of 510 nm.

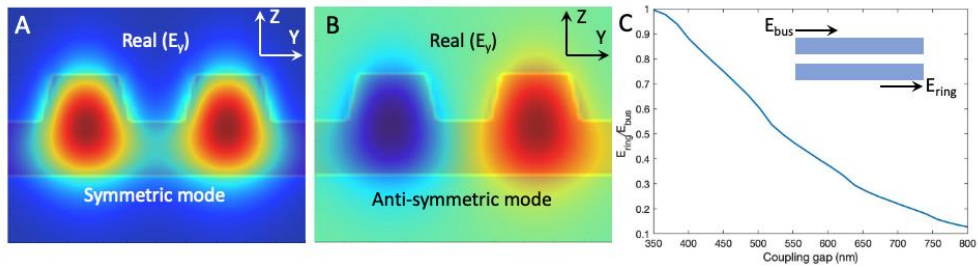

Fig. S1. The E_y component distribution of the symmetric (A) and anti-symmetric (B) mode in the coupled bus and ring waveguides. (C) The simulated value of $E_{ring}/E_{bus}$ as a function of the top coupling gap.

**Section 2. Characterization of the propagation loss of ring resonator**

We estimate the propagation loss of the ring waveguide by measuring the quality factor of a near-critically coupled bare ring resonator with an identical ring waveguide design, fabricated using the same method on a separate undoped thin-film lithium niobate chip. For this bare ring resonator, we modify the coupling region design for a weaker coupling strength: 5 $\mu m$ coupling length, 620 nm coupling gap, to match the small ring loss rate contributed from the low propagation loss. Figure S2 shows the transmission spectrum of a near critically coupled resonance mode of the bare ring resonator (ring length = 878 $\mu m$) at 794.7 nm. We fit the data to a Lorentzian function (see Eq. (1)) and extract the loaded and intrinsic quality factors $Q_L = 5.42 \times 10^5$ and $Q_I = 1.1 \times 10^6$, from which we calculate the propagation loss of 0.7 dB/cm. This value is also comparable with previous results using the same fabrication methods[2]. This propagation loss could be further reduced by depositing a $SiO_2$ cladding layer.

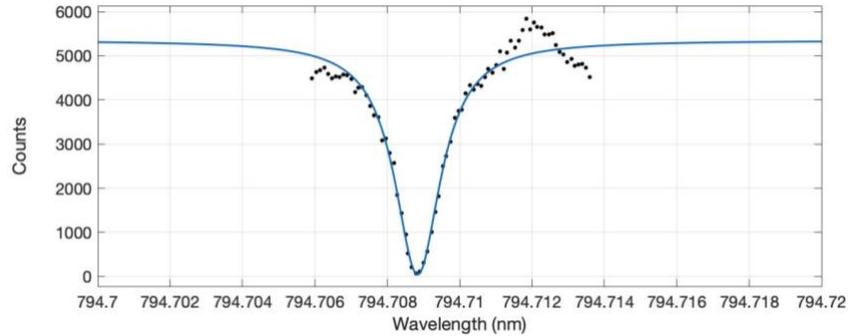

Fig. S2 The transmission spectrum of a near critically coupled bare ring resonator with identical ring waveguide design fabricated on a separate undoped thin-film lithium niobate chip.

**Section 3. Estimation of the power enhancement in the ring waveguide**

The light intensity in the ring waveguide can be much higher than the incident light in the bus waveguide, as the traveling wave in the ring resonator constructively interferes at resonance with the input wave and thus the amplitudes build up. To estimate the intensity enhancement, we consider a simplified model of the ring resonator, as shown in Fig. S3. In the figure, the $E_{i1}$, $E_{i2}$, $E_{t1}$ and $E_{t2}$ represent the two input and output ports at the coupling region, with $r$, $k$ and $a$ denoting the coupling coefficients and amplitude attenuation factor (same definition as in Eq. (1) of the main text). The intensity enhancement factor is given by[3]:

$$\left|\frac{E_{i2}}{E_{i1}}\right|^2 = \left|\frac{ka}{1-ar}\right|^2 \quad (S4)$$

This estimation is utilized as a guidance for the power required to burn the deepest spectral hole in the conventional waveguide, thus we consider the case in which all of the ion absorption is removed by the burning pulse so that $a \approx 1$. This assumption can also ensure an upper bound of burning power used in the waveguide, which would prevent cavity enhancement of the filter extinction ratio to be overestimated. Meanwhile, *r* is extracted to be *r = 0.83* from Fig.2(C). For a lossless coupling process, we calculate *k = 0.56* using the relation $r^2 + k^2 = 1$. With these parameters, we obtain the intensity enhancement in the ring waveguide from Eq.(S1) to be ~10. We note that the geometry difference between the ring and bus waveguide is not considered for this estimation, which could reduce the intra-cavity intensity and power enhancement given that the ring has a wider waveguide. Nonetheless, we expect this will not impact our determination of an upper bound of ~10.

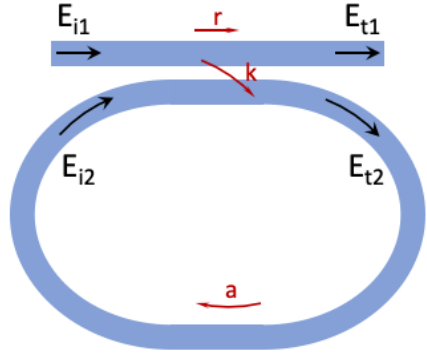

Fig. S3 A simplified model of the ring resonator

**Section 4. Power broadening of spectral hole linewidth and homogeneous linewidth**

The linewidth of a spectral hole increases with the burning power following a power broadening model[4]:

$$\Gamma_{hole} = \Gamma_L \left[ \left(1 + \sqrt{1 + (K/\Gamma_L)^2}\right) \times \left(1 + \sqrt{1 + (K/\Gamma_L)^2 \, e^{-d}}\right) \right]^{1/2}, \quad (S5)$$

where $d$ is the initial ion absorption, $\Gamma_L$ is the fitted homogeneous linewidth that includes linewidth broadening due to laser frequency instability, and $K^2$ is proportional to the excitation power. We fit the waveguide linewidth data (blue line) in Fig. 3C using Eq. (S2), yielding $d = 1.10 \pm 0.12$ and $\Gamma_L = 1.15 \pm 0.21 \, MHz$. We also compare this extracted homogeneous linewidth with the optical coherence time measured using two-pulse photon echo in the same waveguide at 4K, as shown in Fig. S3. The coherence time is extracted from the exponential decay rate as $T_2 = 461 \pm 65 \, ns$, which corresponds to a homogeneous linewidth of $\Gamma_L = 1/\pi T_2 = 0.691 \pm 0.113 \, MHz$. The homogeneous linewidth extracted in both ways are almost consistent. We attribute this shorter coherence time, compared to previous results[5], to a non-ideal sample-stage thermal contact in the cryostat that we use.

At 100 mK, as shown in Fig. 4C in the main text, the photon echo intensity features non-exponential decay because the decoherence is no longer dominated by phonon scattering but by spectral diffusion. By fitting the echo intensity into a Mims' formula: $I = I_0 e^{-2(\frac{2t}{T_2})^x}$, where x represents the spectral diffusion parameter, we extract the coherence time $T_2 = 11.33 \pm 0.28 \, \mu s$ and $x = 1.4$. The homogeneous linewidth is calculated to be $\Gamma_L = 1/\pi T_2 = 28.11 \pm 0.68 \, kHz$, which corresponds to a fundamental limit on the hole linewidth of ~56 kHz.

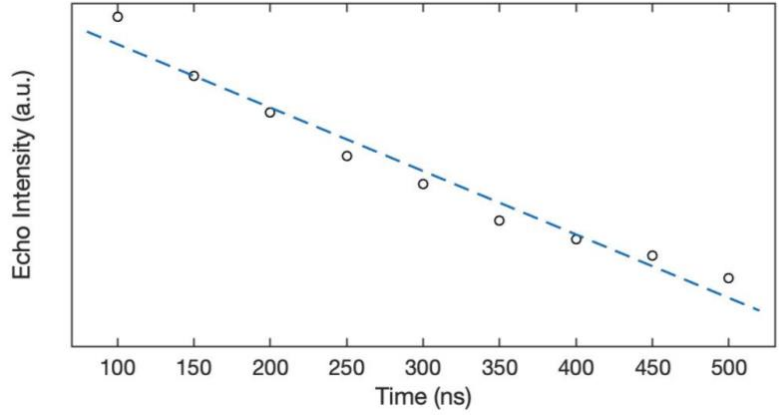

Fig. S4. The echo intensity decays exponentially as a function of delay time in the two-pulse photon echo measurement at 4K.

**Section 5. Characterization of the under-coupled resonance mode**

To characterize the under-coupled resonance mode at 794.13 nm in ring resonator with a 605 nm coupling gap, we measure the transmission spectrum using an attenuated narrowband tunable laser (Msquared). By scanning the wavelength of the laser, we measure the power of the transmitted light as a function of the wavelength, as shown in Fig. S5. We fit the data to the all-pass ring resonator transmission equation (Eq. (1)) and extract $r = 0.935$ and $a = 0.794$. From this, we calculate $\kappa_e = 18.8\ GHz$ and $\kappa_{in} = 64.8\ GHz$ using Eq. (2-3).

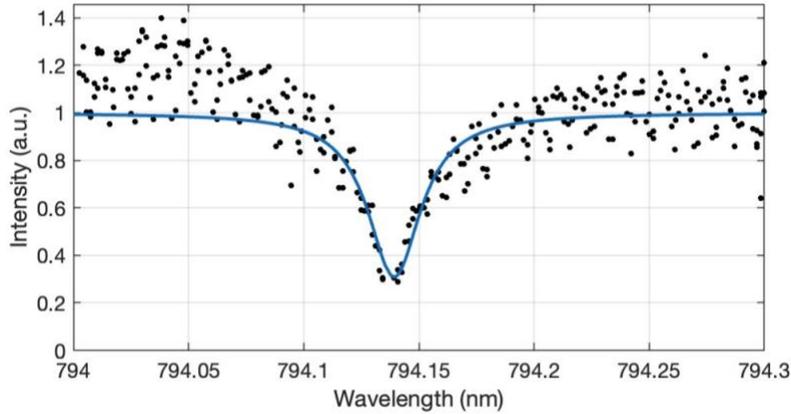

Fig. S5 The transmission spectrum of the under-coupled resonance mode at 794.13 nm

**Section 6. Detuned filter model based on Kramers-Kronig relations**

As discussed in Section 5, the transmission spectrum of the under-coupled ring resonator can be described by Eq. (1):

$$T = \frac{r^2 + a^2 - 2ar\cos(\varphi)}{1 + r^2 a^2 - 2ar\cos(\varphi)} \quad (S6)$$

where the *r* and *a* are extracted to be 0.935 and 0.794, and $\varphi$ is the one-round trip phase given by $\varphi = n\omega L/c$, with *n* representing the real refractive index of the material. We create a detuned bandstop filter by burning a spectral hole at frequency $\omega_1$, detuned by $\Delta\omega$ from the resonant frequency, $\omega_0$. We consider the spectral hole to be Lorentzian shaped with a hole depth of *d* and a linewidth of $\gamma$. This spectral hole modifies the near constant ion absorption coefficient $\alpha_0 = 11.56\ cm^{-1}$ at 794.13 nm into a frequency-dependent function:

$$\alpha(\omega) = \alpha_0 - \gamma^2 d/[(\omega - \omega_1)^2 + \gamma^2] \tag{S7}$$

From this equation, we can obtain the resulting frequency dependent attenuation factor, $a(\omega)$, from absorption coefficient, $\alpha(\omega)$, as:

$$a(\omega) = e^{-(\alpha(\omega)L_a + \alpha_p L)/2}, \tag{S8}$$

where $L_a$ and $L$ are the effective absorption length (length that overlaps with the c-axis) and the total length in the ring, and $\alpha_p$ represents the propagation loss characterized in Section 2. Additionally, this narrow spectral hole will also result in a dispersion effect, which will bring in a frequency-dependent change in real refractive index beyond the original constant index $n_0 = 2.256$. The complex refractive index can be written as $\tilde{n}(\omega) = n(\omega) + ik(\omega)$, where the real part $n(\omega)$ is responsible for the change in the phase $\varphi$, while the imaginary part $k(\omega)$ is related to the absorption coefficient as:

$$k(\omega) = \alpha(\omega)c/2\omega \tag{S9}$$

The $n(\omega)$ is connected to $k(\omega)$ through the Kramers-Kronig relations as:

$$n(\omega) = n_0 + \frac{2}{\pi}\mathcal{P}\int_0^\infty \frac{\omega' k(\omega')}{\omega'^2 - \omega^2} d\omega', \tag{S10}$$

where $\mathcal{P}$ symbolizes the Cauchy principal value. Utilizing this obtained $n(\omega)$, we calculate the phase shift $\varphi(\omega)$ as:

$$\varphi(\omega) = \frac{\omega}{c}[(L - L_a)n_0 + L_a n(\omega)] = \frac{n_0 \omega L}{c} + \frac{2\omega L_a}{c\pi}\mathcal{P}\int_0^\infty \frac{\omega' k(\omega')}{\omega'^2 - \omega^2} d\omega'. \tag{S11}$$

The first term in Eq. S8 represents the phase caused by the original index $n_0$, while the second term denotes the phase resulted from the frequency dependent index change, which only accumulates for an effective length of $L_a$ that aligns with the ion absorption direction. By substituting Eqs. (S4-8) into Eq. (S3), we fit the measured transmission spectrum of the bandstop filters detuned by $\pm 0.006\ nm$ with fitting parameters: $d = 0.3$, $\gamma = 5\ MHz$, as shown in the black curves in Fig. 5E, F of the main text.

**Section 7. Noise analysis of the cavity-enhanced filter for quantum applications**

After the spectral hole burning process, the decay of the residual $^3H_4$ population could create spontaneous emission noise photons at the operation wavelength (794 nm). To evaluate whether this noise would limit our device for quantum information applications, we perform noise analysis under two conditions.

The first condition is under zero magnetic field, we extend the burning pulse from the previously used $50~\mu s$ into $5~ms$ to burn a more persistent hole. This burning pulse ensures sufficient pump duration compared to the $^3H_4$ lifetime ($160~\mu s$), trapping a significant amount of population in the longer-lived $^3F_4$ level. Figure S6 shows the relative extinction ratio (in dB) of the cavity-enhanced filter as a function of time after the burning pulse. The extinction ratio can be fitted into a single exponential decay with a lifetime of $4.5 \pm 0.7~ms$, consistent with the $^3F_4$ lifetime[7].

The blue (orange) curve in Fig. S6 shows the measured (fitted) spontaneous emission noise level, revealing the noise count rate inside the output bus waveguide, as a function of delay time after turning off the 5 ms burning pulse. The measured count rate exponentially decays with a fitted lifetime of $71.2 \pm 3.9~\mu s$. We attribute this faster decay compared to the expected $^3H_4$ lifetime ($160~\mu s$) to a Purcell enhancement[6] enabled by the ring resonator. The count rate decays to ~$1.25 \times 10^3$ counts per second at $380~\mu s$ after the burning pulse, on which time scale the filter extinction ratio only decays by ~ 10%. This noise level corresponds to $2.5 \times 10^{-4}$ counts in the 200 ns window of a 681 kHz input optical pulse bandwidth, which is compatible with our filter, and thus sufficiently small for quantum signal filtering.

The noise level could be further significantly improved by applying a magnetic field. In the second condition, we apply a $150~G$ magnetic field along the c-axis, splitting both the ground and excited states into a pair of hyperfine $Tm^{3+}$ nuclear spin states with a lifetime up to several hours[8]. Exploiting this long-lived hyperfine state, we burn the spectral hole with a 50 ms pulse. With a duration much longer than both $^3H_4$ and $^3F_4$ lifetimes, this burning pulse ensures most of the population is trapped in the hyperfine state. As shown in Fig. S6, there is no measurable degradation for timescales exceeding 10 ms. Longer time scales are not investigated here due to their prolonged measurement times. However, we expect the hole to persist for up to several hours in this condition according to previous works[8]. Compared to the persistent spectral hole, the noise photons decay to unmeasurable levels after 1 ms. Therefore, with a magnetic field we can burn a persistent hole with virtually no noise, enabling quantum operation under a broad range of operating conditions.

It's worth noting that these two different holes burning schemes do not create any degradation in the high extinction ratio of the cavity-enhanced filters demonstrated in Fig. 3, which are created with a shorter burning pulse of $50~\mu s$. For comparison, we generate a spectral hole utilizing a $50~\mu s$ burning pulse (zero magnetic field) and probed after a $10~\mu s$ delay, following the same process as used in Fig. 3, while keeping the other conditions constant to the other two burning schemes (e.g. burning power, temperature, wavelength). As illustrated by the red marker in Fig. S6, the measured extinction ratio is comparable to the values at the same time delay obtained with the other two burning schemes. Therefore, the hole burning schemes could be easily adjusted while maintaining the high extinction ratio.

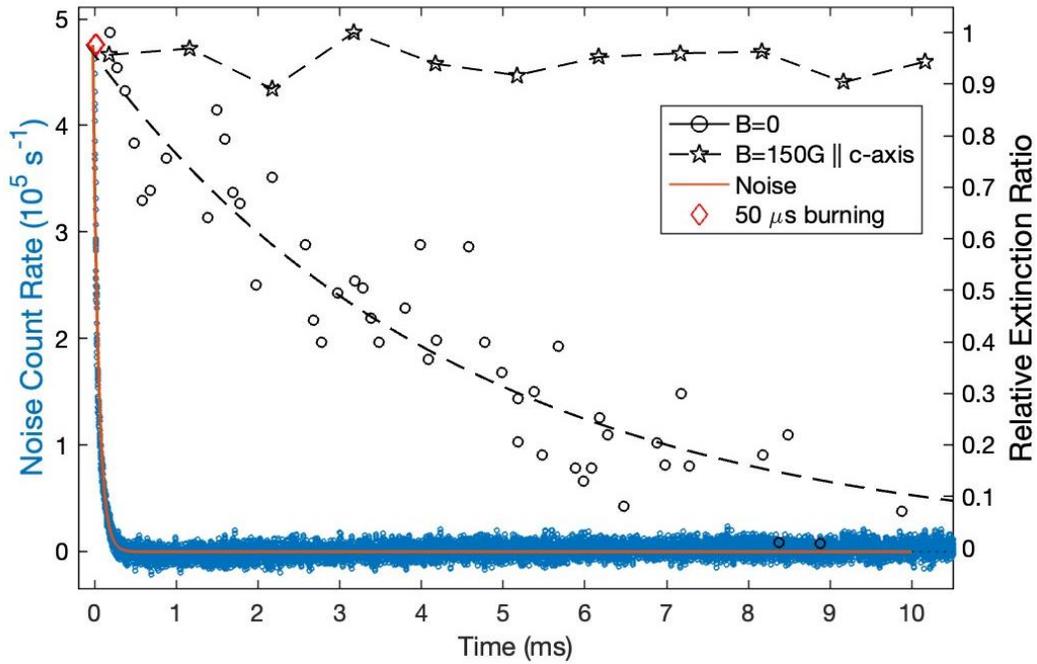

Fig. S6 The dynamics of noise count rate, as well as the extinction ratio of spectral hole created with and without a magnetic field, at 100 mK.

**Section 8. Calculation of the loss of the cavity-enhanced bandpass filter**

From Fig. 2C of the main text, we extracted the self-coupling rate $r = 0.83$ and the one-round amplitude transmission, $a = 0.80$. However, in the measurement of the cavity-enhanced bandpass filter shown in Fig. 3A, the cavity slightly red shifted from 794.535 nm to 794.65 nm (during the measurement, the burning wavelength is set to be 794.65 nm to keep resonant to the shifted cavity). At this red-shifted wavelength, the ion absorption slightly drops (see Fig. 2B, orange line), resulting in a = 0.851. Without any hole burning, the transmission is calculated to be 0.51%, by substituting $r = 0.83$, $a = 0.851$ into Eq. (1). Given the measured spectral hole extinction ratio 110, the center transmission of this bandpass filter is around 56%. Given the measured propagation loss 0.7 dB/cm, we could estimate that within this 2.4 dB background loss, 0.8 dB is from propagation loss ($\kappa_p = 2.1\ GHz$) and 1.6 dB is from remaining ion absorption, which corresponds to $\kappa_p = 4.46\ GHz$.